\documentclass{emulateapj}

\newcommand{\ltsim}{\lower.5ex\hbox{$\; \buildrel < \over \sim \;$}}
\newcommand{\gtsim}{\lower.5ex\hbox{$\; \buildrel > \over \sim \;$}}

\newcommand{\ha}{H$\rm\alpha$}
\newcommand{\hb}{H$\rm\beta$}

\begin{document}

\title{Evidence for Line Broadening by Electron Scattering
in the Broad Line Region of NGC 4395}

\author{Ari Laor}
\affil{Physics Department, Technion, Haifa 32000, Israel}
\email{laor@physics.technion.ac.il}

\begin{abstract}
A high quality Keck spectrum of the \ha\ line in NGC~4395 reveals symmetric exponential
wings, $f_v\propto e^{-v/\sigma}$, with $\sigma\simeq 500$~km~s$^{-1}$. The wings extend
out to $\gtsim 2500$~km~s$^{-1}$ from the line core, and down to a flux density
of $\ltsim 10^{-3}$ of the peak flux density. Numerical and analytic calculations indicate
that exponential wings are
expected for optically thin, isotropic, thermal electron scattering. Such scattering
produces exponential wings with
$\sigma\simeq 1.1\sigma_e(\ln\tau_e^{-1})^{0.45}$, where $\sigma_e$ is the electron
velocity dispersion, and $\tau_e$ is the electron scattering optical depth.
The \ha\ wings in NGC~4395 are well fit by an electron scattering model with $\tau_e=0.34$,
and an electron temperature $T_e=1.1\times 10^4$K. Such conditions are
produced in photoionized gas with an ionization parameter $U\simeq 0.3$, as expected
in the broad line region (BLR). Similar analysis of the
[O~III]~$\lambda 5007$ line yields $\tau_e<0.01$, consistent with the lower ionization
 in the narrow line region.
If the electron scattering interpretation is correct, there should be a
tight correlation between $\tau_e$ and the ionizing flux on time scales
shorter than the BLR dynamical time, or $\sim 1$~week for NGC~4395.
In contrast, the value of $\sigma$ should remain nearly constant on
these time scales. Such wings
may be discernible in other objects with unusually narrow Balmer lines, and they
can provide a useful direct probe of $T_e$ and $\tau_e$ in the BLR.

\end{abstract}


\section{Introduction}
Early studies of the broad line profiles in Type~1 Active Galactic Nuclei (AGN)
suggested that the large observed line widths ($\gtsim 3000$~km~s$^{-1}$) are produced by
electron scattering within the photoionized gas
(Weymann 1970; Mathis 1970). With the development of photoionization
models for the Broad Line Region (BLR), it was however realized that the expected
electron scattering optical depth, $\tau_e$, within the BLR gas is $\ll 1$,
which is too low to
explain the observed line profiles (Davidson \& Netzer 1979).
Later studies of the size of the BLR, based on reverberation mappings,
indicated that the BLR is more compact than previously thought (e.g. Rees,
Netzer, \& Ferland 1989),
and thus $\tau_e$ is expected to be higher than previously thought, though it
is not expected to
reach $\tau_e> 1$, required to explain the line profiles purely by
electron scattering. There is also good evidence now that the line broadening
in the BLR is mostly due to virialized bulk motion (e.g. Peterson \& Wandel 2000).
However, electron scattering by a significantly hotter gas outside the BLR gas
($T\sim 10^{6-7}K$ vs. $T\sim 10^4K$) remains a viable option (Shields \& McKee 1981;
Kallman \& Krolik 1986;
Emmering et al. 1992; Bottorff et al. 1997). Evidence for electron scattering
of continuum photons in AGN is provided by wavelength independent polarization
in some objects
(e.g. Antonucci et al. 1993), but there is yet no conclusive detection of electron
scattering effects on the broad line profiles.

A possible exception may occur in NGC~1068, where spatially resolved
spectropolarimetry reveals a broadened \ha\ line, which may result from
electron scattering in warm ($T\sim 10^5K$) gas (Miller, Goodrich \& Mathews 1991).
A spectropolarimetric search for polarized broad lines in low luminosity AGN
(Barth et al. 1999) revealed a polarized BLR component in the lowest luminosity type~I
AGN, NGC~4395.
Based on that detection, Barth et al. suggested that higher S/N spectropolarimetry
may reveal a broadened \ha\ component, if the polarization is induced by electron
scattering. Here we report on a high quality Keck observation of the \ha\ line
in NGC~4395, which indeed reveals extended symmetric exponential wings (though their
polarization is unknown). As
shown below, these wings are
well fit by optically thin electron scattering by electrons at
$T_e=1.14\times10^4K$, with $\tau_e=0.34$, which are plausible conditions for
the BLR gas.

The paper is structured as follows. Section 2 describes the observations
which revealed the exponential wings in the \ha\ profile of NGC~4395. Section 3
describes the numerical calculation of the electron scattering line profiles, which yield
exponential wings, $f_v=f_0e^{-v/\sigma}$. A fitting formula for $\sigma$
as a function of $T_e$ and $\tau_e$ is provided,
together with a simple analytic derivation.
In section 4 the observed \ha\ profile is fit with the electron
scattering model, yielding $T_e$ and $\tau_e$, the origin of the
scattering gas is discussed, and some predictions are made. The main conclusions are
summarized in section 5.

\section{Observations}

 NGC~4395 was observed with the Keck-II 10~m telescope using the
Echellette Spectrograph and Imager (ESI; Sheinis et al. 2002); a slit
width of $0^{``}.5$, and a spectral
resolution of $\lambda/\Delta\lambda=8,000$, or 0.820~\AA\ near \ha,
sampled at 0.259~\AA\ per pixel. Two 20 minute exposures were made on 2002
December 2 UT, yielding a S/N ranging from 50 per pixel at the continuum near
\ha\ to 400 at the line peak (for the reduction details see Laor et al. 2006).

The continuum level is estimated using a linear fit to the observed flux
densities of $4.32\times 10^{-16}$~erg~s$^{-1}$~cm$^{-2}$~\AA$^{-1}$ at
observed frame 6480~\AA, and $4.20\times 10^{-16}$~erg~s$^{-1}$~cm$^{-2}$~\AA$^{-1}$
at 6640~\AA. The implied spectral slope of $f_{\lambda}\propto \lambda^{-1.15}$
between 6480~\AA\ and 6640~\AA\ is consistent with the global optical continuum
spectral slope in the ESI spectrum, although the flux at 6640~\AA\ appears to
be somewhat blended with the extended blue wing of the He~I~$\lambda 6678.15$ line.
We estimate
the continuum placement uncertainty as
$\sim 3\times 10^{-18}$~erg~s$^{-1}$~cm$^{-2}$~\AA$^{-1}$. This uncertainty implies
a potential systematic error of $\sim 10$\% in the net line flux
at $|v|=2500$~km~s$^{-1}$, increasing to
$\sim 100$\% at $|v|=3000$~km~s$^{-1}$. We therefore consider
the \ha\ line profile reliable at $|v|<2500$~km~s$^{-1}$, where any systematic
flux error is $<10$\%. The continuum 1~$\sigma$ noise level per pixel is
$9.5\times 10^{-18}$~erg~s$^{-1}$~cm$^{-2}$, and thus the statistical uncertainty
per pixel is $\sim 30$\% at $|v|=2500$~km~s$^{-1}$. We find a S/N$>10$ per resolution element
at $|v|<2000$~km~s$^{-1}$, increasing to $>70$
at $|v|<1000$~km~s$^{-1}$. Overall, the high quality of the ESI spectrum allows us
to reliably probe the \ha\ profile over an unprecedented range of a factor of 1000 in
flux density, from $3.58\times 10^{-14}$~erg~s$^{-1}$~cm$^{-2}$~\AA$^{-1}$ at $v=0$~km~s$^{-1}$
to $\sim 3\times 10^{-17}$~erg~s$^{-1}$~cm$^{-2}$~\AA$^{-1}$ at $|v|=2500$~km~s$^{-1}$.

Figure 1, upper panel, presents the overall \ha\ line profile at $|v|<3000$~km~s$^{-1}$.
The nearly pure linear shape in log $f_E$ versus $v$ is very prominent, implying exponential
wings. The middle panel demonstrates the asymmetry present in the $|v|<1000$~km~s$^{-1}$
region of the broad line profile, with excess blue wing flux. Note that the wings at
$|v|>1000$~km~s$^{-1}$ show an opposite asymmetry (see upper panel), however
there is a very good match of the red and blue wings when the spectrum
is reflected with respect to the $v=50$~km~s$^{-1}$ point (lower panel). The two
wings are well matched by an exponential
profile, $f_E\propto e^{-v/\sigma}$, with $\sigma=500$~km~s$^{-1}$. These pure symmetric
exponential wings suggest a scattering origin. Below we show that exponential
wings are indeed a generic prediction of optically thin isotropic
electron scattering.

\begin{figure}
\includegraphics[angle=0,scale=.45]{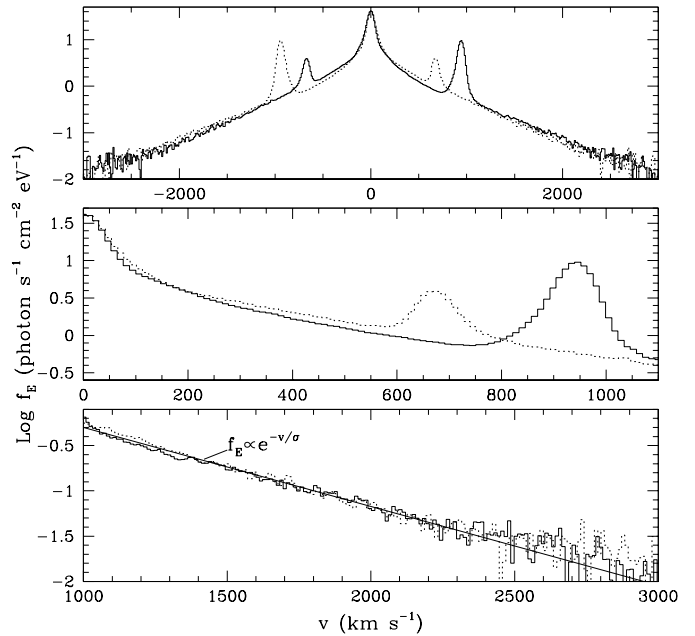}
\caption{Upper panel: The observed \ha\ profile in NGC~4395 (solid
histogram), and the reflected profile (dotted histogram). Note that
the blue wing is higher than the red wing at $|v|<1000$~km~s$^{-1}$,
while the reverse is true at $|v|>1000$~km~s$^{-1}$. Middle panel:
An expanded view of the line core. The narrow \ha\ line dominates
at $|v|<150$~km~s$^{-1}$, and the narrow [N~II] lines dominate
around $v\sim 950$~km~s$^{-1}$, and $v\sim -670$~km~s$^{-1}$.
Lower panel: An expanded view of the extended wings at
$1000<|v|<3000$~km~s$^{-1}$. The velocity scale in this panel
was shifted from $v$ to $v-50$~km~s$^{-1}$. Note the remarkable
symmetry of the wings in the shifted scale, and their nearly pure exponential
form. The solid line is an exponential function with
$\sigma=500$~km~s$^{-1}$.}
\end{figure}

\section{Calculations}
\subsection{Numerical}
A complete calculation of electron scattering line profiles involves a calculation of
the photon diffusion simultaneously in both real space and in energy space. Such a calculation
can be carried out using a Monte Carlo approach, which is straightforward to
implement, though the finite number of photons used 
tends to produce non smooth curves (e.g. Lee 1999) compared to analytic solutions.
Analytic solutions
for the local diffusion in space, described by the radiative transfer equation,
and diffusion in energy, described by the energy redistribution function, have been
derived and
solved numerically using various simplifying assumptions (e.g. Chandrasekhar 1960;
Mihalas 1978; and references therein). Here we make two major simplifying assumptions.
First, the electron scattering gas is assumed to be optically thin, thus a detailed
radiative transfer solution is not required. This simplification is justified by
the observation that the dominant effect on the line broadening in the BLR appears to
be gravity dominated bulk motion, rather than electron scattering. Gravity is 
indicated by the good correspondence between
black hole mass estimates assuming a virialized BLR and mass estimates based on
the host bulge properties (Laor 1998; Ferrarese et al. 2001).
Second, we assume an isotropic electron velocity distribution, and an isotropic 
illumination of the scattering electrons, which is likely to be a good approximation
for the integrated scattered radiation.
 These assumptions
lead to the following remarkably simple energy redistribution function,
derived by Rybicki \& Lightman (1979, hereafter RL79. Eq. 7.24 there),

\begin{equation}
j(x,\beta)=\frac{1-\beta^2}{4\beta^2}[(1+\beta)x-(1-\beta)]\ \ \ \ {\rm for}\ \  \frac{1-\beta}{1+\beta}<x<1
\end{equation}
\begin{equation}
j(x,\beta)=\frac{1-\beta^2}{4\beta^2}[(1+\beta)-x(1-\beta)]\ \ \ \  {\rm for}\ \ 1<x<\frac{1+\beta}{1-\beta},
\end{equation}
\begin{equation}
j(x,\beta)=0\ \ \ \  {\rm for}\ \ x>\frac{1+\beta}{1-\beta}\ \ \ \  {\rm or}\ \ x<\frac{1-\beta}{1+\beta},
\end{equation}
where $x\equiv e/e_0$, and $e_0,e$ are the observed photon energy before and after scattering,
$v=\beta c$ is the scattering electron velocity, and $j(x,\beta)$ is the number of photons scattered
per $dx$, which obeys the photon conservation $\int_0^{\infty}j(x,\beta)dx=1$.  The expressions
above for $j(x,\beta)$ apply for elastic scattering in the electron rest frame, which is a very
good approximation for the applications below, where $\beta\ll 1$ and $h\nu\ll m_ec^2$.

To obtain the redistribution function for a thermal electron distribution, $j_t(x)$, one needs
to integrate the above expression for $j(x,\beta)$ over a thermal velocity distribution
function. Thus
\begin{equation}
j_t(x)=\frac{4}{b^3\sqrt{\pi}}\int_0^{\infty}j(x,\beta)v^2e^{-(v/b)^2}dv
\end{equation}
where $b\equiv 2kT_e/m_e$, and $T_e$ is the electron temperature.
Figure 2 (upper panel) shows $j(x,\beta)$ and $j_t(x)$. Note that
despite the thermal averaging, $j_t(x)$ remains sharply peaked.

\begin{figure}
\includegraphics[angle=0,scale=.45]{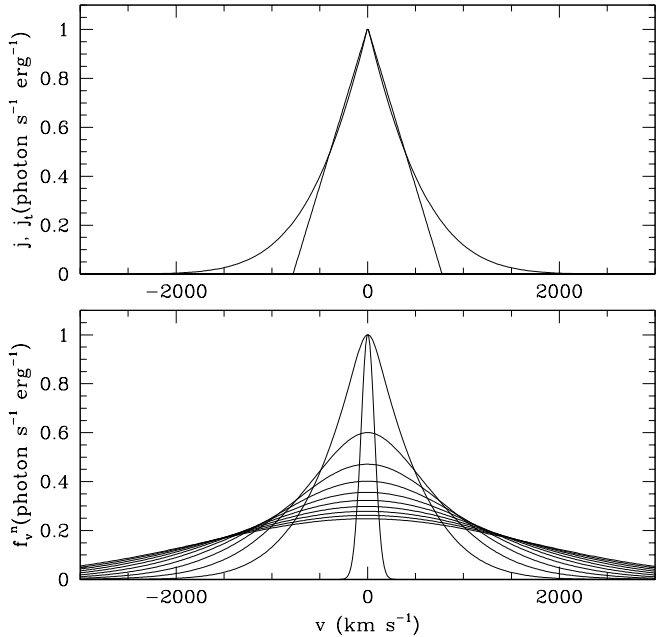}
\caption{Upper panel: The single velocity redistribution function $j(x,\beta)$
for $T_e=10^4$~K (triangle shaped curve), and the thermal averaged function
$j_t(x)$, scaled to match $j(x,\beta)$ at $x=1$ (i.e. at $v=0$~km~s$^{-1}$).
Note that despite the thermal averaging the core of the redistribution function
is almost unchanged. Lower panel: The effects of repeated scattering. The
narrow Gaussian is the unscattered line profile, and the broader curves
correspond to $f_v^n$ with $n=1-10$. The number of photons (area under each
curve) is assumed to be conserved, and the unscattered line profile was
scaled to match the peak of the $n=1$ curve.}
\end{figure}

Since the scattering medium is optically thin, each scattering redistributes
the observed line profile through $j_t(x)$, and reduces the number of
photons by a factor of $\tau_e$. Thus, the line profile following
$n+1$ scattering is given by a convolution of $j_t(x)$
with the $n$ times scattered profile, i.e.
\begin{equation}
f_v^{(n+1)}=\tau_e\int_0^{\infty} f_{v'}^nj_t[(v-v')/c]dv'
\end{equation}
where $f_v^n$ is the profile of a line scattered $n$ times. Since
$\beta\ll 1$, a simple Doppler relation $de/e_0=dv/c$ can be assumed.
Figure 2 (lower panel) shows $f_v^0$, an arbitrary seed line
profile, and $f_v^n$ for $n=1-10$, illustrating how the
line gets broader and shallower with repeated scattering by a given
electron population. The total scattered line profile is then
given by
\begin{equation}
f_v^{\rm scat}=\sum_{n=1}^{\infty}f_v^n.
\end{equation}
Figure 3 shows the
results of numerical calculations of $f_v^{\rm scat}$ for a range of
$T_e$ and $\tau_e$ values. As both $T_e$ and $\tau_e$ increase, the line profile gets
broader, with a relatively weak dependence of the width on $\tau_e$ and a stronger
dependence on $T_e$. Note that in all cases $\log f_v^{\rm scat}$ approaches a
linear function of $v$ in the extended wings, or equivalently
\begin{equation}
f_v^{\rm scat}=f_0e^{-v/\sigma}.
\end{equation}
Figure 4 shows a plot
of $\sigma$, measured from the numerical calculations, as a function of
$\tau_e$ and $T_e$. The solid lines is a fit to the numerical
results. The fitting function has the simple form,
\begin{equation}
\sigma=1.1\sigma_e(\ln\tau_e^{-1})^{-0.45},
\end{equation}
where $\sigma_e$ is the thermal electron velocity dispersion,
\begin{equation}
\sigma_e=\sqrt{kT_e/m_e}=389.2T_4^{1/2}~{\rm km~s}^{-1},
\end{equation}
and $T_4\equiv T_e/10^4$.
This fitting function is accurate to 1.5\% in the range $0.01\le\tau_e\le0.8$.

\begin{figure}
\includegraphics[angle=0,scale=.45]{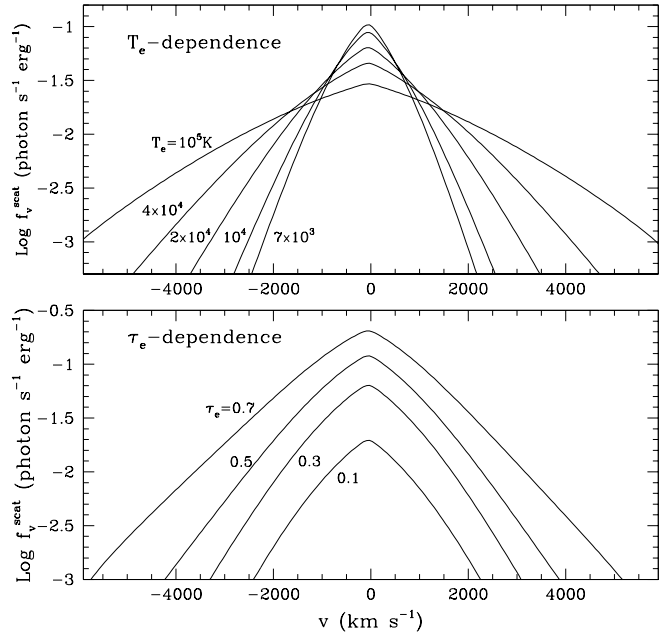}
\caption{The dependence of $f_v^{\rm scat}$ on $T_e$ and $\tau_e$.
Upper panel: The dependence on $T_e$ for $\tau_e=0.3$. The line wings approach
a linear shape in log $f_v^{\rm scat}$ versus $v$, i.e. an exponential form
$f_v^{\rm scat}\propto e^{-v/\sigma}$, at high $v$. Note that $\sigma$ increases
significantly with $T_e$.
Lower panel: The dependence on $\tau_e$ for $T_e=2\times 10^4$~K. The integrated
scattered flux is proportional to $\tau_e$, however $\sigma$ is only
weakly dependent on $\tau_e$.}
\end{figure}

\subsection{Analytic Derivations}

\subsubsection{An Oversimplified Derivation}
One may attempt to understand the above fit equation based on the
following very simple derivation. This derivation is actually
oversimplified, and the resulting expression for $\sigma$ lacks accuracy,
but it provides insight for understanding the basic
reason for the formation of exponential profile wings.

This derivation follows closely the derivation of the power law spectral
form produced through optically thin electron scattering by relativistic
electrons, as described in RL79 (section 7.5 there).
Here we have optically thin electron scattering by non relativistic
electrons. Let $\sigma_e$ be the mean electron velocity. The energy
amplification of a photon scattered at right angle is
\begin{equation}
e_1/e_0\simeq 1+\sigma_e/c.
\end{equation}
After $n$ such scattering the photon energy will be
\begin{equation}
e_n\simeq e_0(1+\sigma_e/c)^n, \ \ {\rm or}\ \ e_n\simeq e_0(1+n\sigma_e/c) ,
\end{equation}
where the last step is valid for
$n\sigma_e/c\ll 1$.
Changing to velocity scale
\begin{equation}
v_n\equiv (e_n/e_0-1)c, \ \ \
{\rm gives}\ \ v_n\simeq n\sigma_e.
\end{equation}
The fraction of photons scattered $n$ times
is $\tau_e^n$, and thus
\begin{equation}
f(v_n)\simeq f(0)\tau_e^n,\ \ \ {\rm or}\ \
f(v_n)\simeq f(0)\tau_e^{v_n/\sigma_e}.
\end{equation}
This is equivalent to
\begin{equation}
f(v_n)\simeq f(0)e^{\ln \tau_e v_n/\sigma_e},
\end{equation}
i.e. an exponential line
profile, where
\begin{equation}
f(v)\propto f(0)e^{-v/\sigma}\ \ \ {\rm and}\ \ \sigma=\sigma_e/\ln \tau_e^{-1}.
\end{equation}
The crucial difference here from relativistic electron scattering is that the
energy increment with repeated scattering is a constant linear factor, rather
than a constant multiplicative factor, which would lead to a power law rather than
an exponential form.

The above derivation of $\sigma$ does not yield the correct dependence on
$\ln \tau_e$. The oversimplification in the above derivation is in the
assumption that the energy redistribution function can be approximated
as a $\delta$ function at $e_1/e_0\simeq 1+\sigma_e/c$, while as Eqs.1-3
above show, the photons actually get redistributed nearly symmetrically from
$1-\sigma_e/c$ to $1+\sigma_e/c$. For highly relativistic electrons the
strong Doppler beaming produces a highly asymmetric and strongly peaked
redistribution function (at $e_1/e_0\simeq\gamma^2$), and the simplified derivation of the
power law spectral slope through repeated scattering (RL79) is adequate.

\begin{figure}
\includegraphics[angle=0,scale=.45]{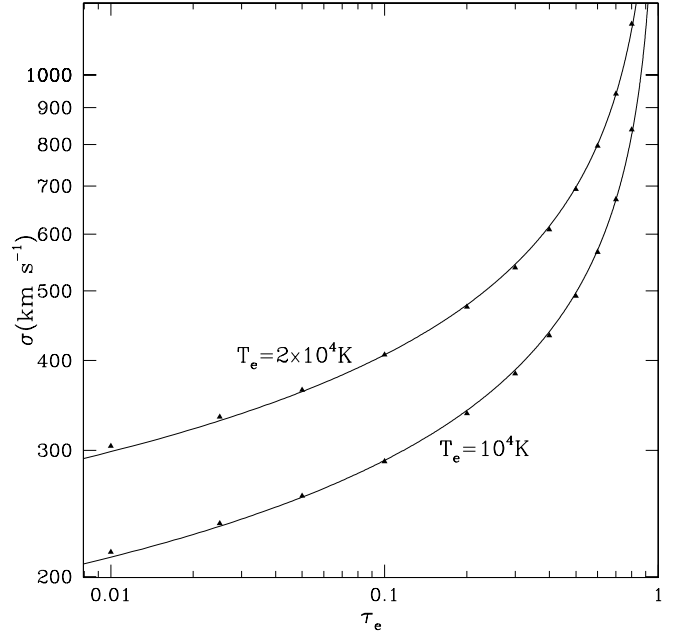}
\caption{The dependence of $\sigma$ on $\tau_e$ for two values of
$T_e$. The filled triangles are values measured from numerical
calculations of $f_v^{\rm scat}$, and the solid lines are analytic
fits of the form $\sigma=1.1\sigma_e(\ln\tau_e^{-1})^{-0.45}$
to the numerical results (see text). This fit is accurate to
1.5\% in the displayed range.}
\end{figure} 

\subsubsection{A Simplified Derivation}

Taking into account the shape of the redistribution function leads to a more accurate
dependence of $\sigma$ on $\tau_e$, as described below.

The thermal redistribution function can be approximated as a Gaussian
with the velocity dispersion of the scattering electrons,
$j_t(x)\propto e^{-v^2/b^2}$. The
seed line profile is assumed to have a smaller velocity width than that
of the scattering electrons, so it can be conveniently approximated as a $\delta$ function,
$f^0_v=\delta(v)$. Thus, the first scattering (Eq.5) yields $f^1_v=j_t(x)$,
and repeated scattering amount to repeated convolutions with
$j_t(x)$. Such convolutions yield Gaussians with a velocity dispersion which grows as
$\sqrt{n}$, i.e. we can use the approximation
\begin{equation}
f_v^n=K\frac{\tau_e^n}{\sqrt{n}}e^{-v^2/nb^2},
\end{equation}
where $K$ is an arbitrary constant, and the $1/\sqrt{n}$ term comes from the
Gaussian normalization. The total
scattered line profile is then given by
\begin{equation}
f_v^{\rm scat}=K\sum_{n=1}^{\infty}\frac{\tau_e^n}{\sqrt{n}}e^{-v^2/nb^2}.
\end{equation}
Unfortunately, there is no simple analytic solution for the sum of this
series. Note, however, that only a few terms in the above series contribute
at any given $v$. The contribution of terms with $n\ll v^2/b^2$ is small because of the
large negative exponential factor, while the contribution of terms with
$n\gg v^2/b^2$ is reduced by the  $\tau_e^n/\sqrt{n}$ factor. The maximum contribution
to $f_v^{\rm scat}$ in the above sum is provided by the term $n_{\rm max}$,
which is obtained from the requirement
\begin{equation}
\frac{df_v^n}{dn}=(\ln \tau_e-\frac{1}{2n}+
\frac{v^2}{n^2b^2})f_v^n=0,
\end{equation}
which holds at $n=n_{\rm max}$. Since $f_v^n\neq 0$ one gets
a quadratic equation for $n_{\rm max}$
\begin{equation}
n_{\rm max}^2-\frac{n_{\rm max}}{2\ln\tau_e}+\frac{v^2}{b^2\ln\tau_e}=0 .
\end{equation}
The second term can be neglected compared to the first and third terms
when $n>1$ and $\tau_e<0.6$. Thus, the solution is
\begin{equation}
n_{\rm max}=\frac{v}{b\sqrt{\ln\tau_e^{-1}}}.
\end{equation}
Substituting $n_{\rm max}$ in the general expression for $f_v^n$ gives
\begin{equation}
f_v^{n_{\rm max}}=K\frac{e^{-2v\sqrt{\ln\tau_e^{-1}}/b}}{\sqrt{v/(b\sqrt{\ln\tau_e^{-1}})}}.
\end{equation}
We further show that $f_v^n$ indeed has a narrow peak around $n=n_{\rm max}$.
We define the peak width, $\Delta n$, through the relation
$f_v^{n_{\rm max}+\Delta n}/f_v^{n_{\rm max}}=0.5$, which yields
after some steps the condition
\begin{equation}
\tau_e^{2(\Delta n)^2/n}\simeq 0.5,
\end{equation}
or
\begin{equation}
\Delta n \simeq 0.6n^{1/2}/ \sqrt{\ln\tau_e^{-1}},
\end{equation}
i.e. only the nearest
1-2 bins to $n_{\rm max}$ contribute significantly for $n\ltsim 10$. Thus, at each $v$ one
can approximate the above sum over $n$ by just a single term at
$n=n_{\rm max}$, i.e. $f_v^{\rm scat}=f_v^{n_{\rm max}}$. To verify that
$f_v^{\rm scat}$ has the correct exponential form we calculate
\begin{equation}
\frac{1}{\sigma}=\frac{d \ln f_v^{\rm scat}}{dv}=\frac{2\sqrt{\ln\tau_e^{-1}}}{b}
-\frac{1}{2v} .
\end{equation}
Thus, a pure exponential line profile, i.e. $\sigma$ independent of $v$,
is obtained when  the second term on the right is negligible,
which occurs at $v>b/4\sqrt{\ln\tau_e^{-1}}$. The exponential slope is
$\sigma=b/2\sqrt{\ln\tau_e^{-1}}$, or $\sigma=0.71\sigma_e\sqrt{\ln\tau_e^{-1}}$,
which matches well the $\ln\tau_e$ dependence of the fitting function to the
numerical solution (Eq.8).

\section{Results and Discussion}

\subsection{The best fit electron scattering model}
To model the shape of the electron scattering wings one needs an estimate of the
seed line profile. As shown below, there is no evidence for detectable electron
scattering wings for lines originating in the narrow line region (NLR), and we
therefore use only the broad component of \ha\ for the seed profile. To obtain
the BLR \ha\ component, we
subtract off the narrow \ha\ component and the narrow [N~II] lines from the observed
\ha\ profile. The broad component underlying the narrow lines is interpolated
using a spline fit, where we use the minimum number of spline
points required to get both an acceptable $\chi^2$, and also matching profiles for the
narrow [N~II] doublet lines (Laor et al. 2006).

The observed broad line core at $|v|<1000$~km~s$^{-1}$ shows a noticeable asymmetry,
with up to $\sim 50$\% higher flux in the blue wing (Fig.1, middle panel). This appears
to be a persistent effect in NGC~4395, as it is also present in high resolution Keck spectra
taken about a year and eight years earlier (Fig.1 in Laor et al. 2006). In contrast, the
wings at $|v|>1000$~km~s$^{-1}$ are generally symmetric to $<10$\% (with respect to
$v=50$~km~s$^{-1}$, Fig.1, lower panel). Since our electron scattering model generates
symmetric line profiles, one must assume that the observed asymmetric core profile
represents the unscattered seed line. To estimate the shape of
the seed line wings, the observed profile is multiplied by a weight function of the
arbitrary functional form $e^{-v^2/w^2}$, which produces a smooth cutoff in the the
wings at $|v|>w$. We find that $w=600$~km~s$^{-1}$ at
$v>0$~km~s$^{-1}$, and $w=800$~km~s$^{-1}$ at $v<0$~km~s$^{-1}$, are about
the minimum values required to exclude the inner asymmetric core emission, and
allow the electron scattering model to fit only the extended symmetric wings. The exact
value of $w$ cannot be accurately constrained, and the true seed profile may
be broader than assumed here. The seed profile is further multiplied
by $e^{-\tau_e}$, which represents the wavelength independent suppression due to
the electron scattering, where $\tau_e$ is determined below.

The seed line profile obtained above is now used as the input line profile for
our electron scattering code. The model free parameters are $T_e$ and $\tau_e$,
and we iterate over different values for these parameters 
until a satisfactory fit is obtained between the
observed broad \ha\ profile, and the sum of the scattered + seed line profiles
(see Figure 5).
The scattering medium is assumed to have a bulk velocity of $v=50$~km~s$^{-1}$, with
respect to the narrow line peak, as suggested by the observed asymmetry of the
extended wings. The best fit is obtained for $T_e=1.14\times 10^4$~K, and
$\tau_e=0.34$, which yields $\chi^2=282$ for 243 degrees of freedom, where
$\chi^2$ is calculated for $|v|<2500$~km~s$^{-1}$, excluding the line core
region, $-900<v<1200$~km~s$^{-1}$, affected by the narrow lines. The
best fit $T_e$ and $\tau_e$ depend on the assumed shape for the seed line
profile. Increasing the wings cutoff velocity, $w$, reduces the best fit $\tau_e$
and increases $T_e$. For example, adopting $w=1200$~km~s$^{-1}$ yields a best
fit solution with $T_e=1.7\times 10^4$~K, and $\tau_e=0.17$. However, at such a
high $w$ the seed line contribution becomes negligible ($<20$\% of total line flux)
at $|v|\gtsim 1500$~km~s$^{-1}$, compared to $|v|\gtsim 1000$~km~s$^{-1}$
for the minimum $w$ solution. The observed wings are well fit by a featureless
exponential profile already at $|v|\gtsim 1000$~km~s$^{-1}$ (Fig.1, lower panel),
which suggests this region is dominated by a single component, as we find for the
minimum $w$ solution. We therefore adopt below the best fit minimum $w$ solution,
although one cannot rule out a higher $w$ solution, where the seed line profile
and the scattered wings conspire to produce a single featureless exponential
profile at $|v|\gtsim 1000$~km~s$^{-1}$.

Once the seed line profile is set, there is essentially no degeneracy in the solution,
as $\tau_e$ is fixed uniquely by the ratio of seed to scattered photon flux ratio, and
$T_e$ is then fixed by $\tau_e$ and the observed exponential slope, $\sigma$. The formal
uncertainties in $T_e$ and $\tau_e$ due to the available S/N here
are $\le 3-5$\%. However, our assumption of an isothermal and isotropic scattering
medium may be a significant oversimplification, and systematic errors
$\gtsim 10$\% are quite possible due to our simplified modeling.

\begin{figure}
\includegraphics[angle=270,scale=.34]{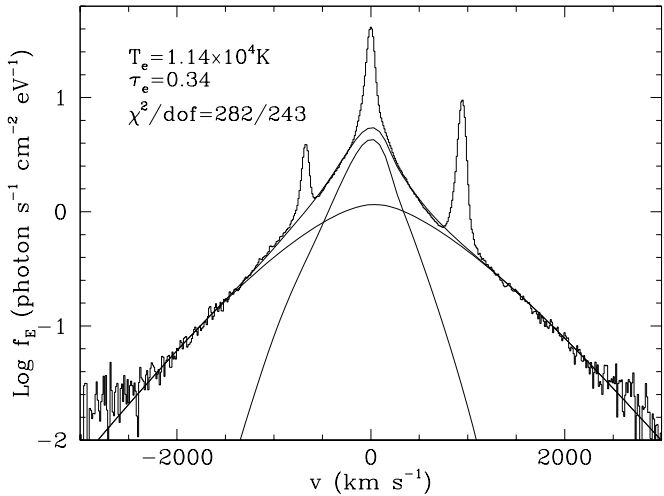}
\caption{The observed \ha\ profile (solid histogram), and the
best fit optically thin electron scattering model (solid line).
The input seed line profile and the electron scattered profile are also
shown. The best fit parameters and the fit $\chi^2$ are indicated on the
plot. }
\end{figure}

\subsection{Where does the scattering takes place?}
The best fit $T_e$ is typical of photoionized gas in the BLR or the NLR. To
clarify whether the scattering gas resides in the BLR or in the NLR, we
looked for evidence for electron scattering wings in the profile of the
[O~III]~$\lambda 5007$ line, the strongest narrow line in the spectrum of
NGC~4395. Figure 6 shows the continuum subtracted line profile. Weak broad
wings appear to be present at $|v|>500$~km~s$^{-1}$. To determine the implied
$\tau_e$, one needs an estimate of the seed line profile.
As for the \ha\ line, we multiplied the observed profile by a weight
function of the form $e^{-v^2/b^2}$, using here $b=400$~km~s$^{-1}$.
We fixed $T_e$ at $1.14\times 10^4$~K, the value measured for \ha, and obtained a reasonable
fit to the wing profile using $\tau_e=0.01$.
However, weak broad wings are also present in the instrumental line
spread function (LSF) at a level of $\sim 10^{-3}$ of the peak line flux
density. Convolving the seed [O~III]~$\lambda 5007$ line profile with
the instrumental LSF produces wings which are also roughly consistent with
the observed wing flux (Fig.6). We thus conclude that most, or possibly all,
of the [O~III]~$\lambda 5007$ wing flux may be instrumental, and therefore
one can only put an upper limit
of $\tau_e<0.01$ on a $T_e\simeq 10^4$~K scattering medium at, or
outside, the NLR. 

\begin{figure}
\includegraphics[angle=270,scale=.34]{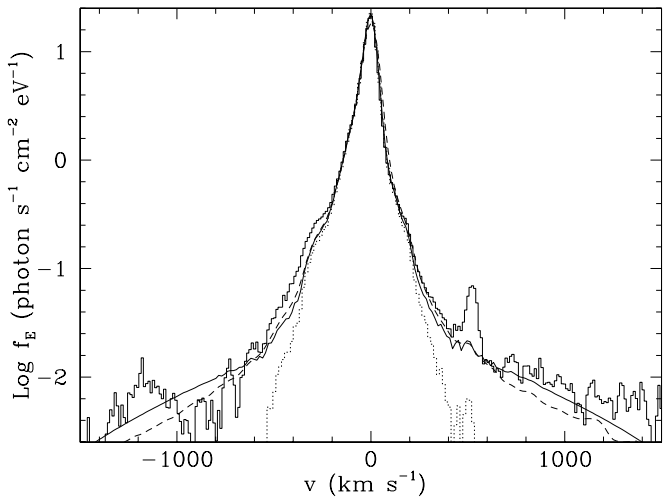}
\caption{The observed [O~III]~$\lambda 5007$ line profile (solid histogram),
the assumed seed line profile (dotted histogram)), and
the best fit optically thin electron scattering model (solid line) having $\tau_e=0.01$
and $T_e$ fixed at $1.14\times 10^4$~K. The wings are also
consistent with instrumental broadening (dashed line), and therefore we deduce that
$\tau_e<0.01$.
The feature at $v\sim 500$~km~s$^{-1}$ is the He~I~$\lambda 5015.68$ line.}
\end{figure}

In contrast to the [O~III]~$\lambda 5007$ line, the
instrumental LSF has a negligible effect on the \ha\ line wings. For example,
the instrumental LSF, acting on the narrow \ha\ component, produces a flux density 
at $|v|=1500$~km~s$^{-1}$ which is $<10^{-4}$ of the peak flux density (Fig.6),
while the observed \ha\ flux density at $|v|=1500$~km~s$^{-1}$ is
$\sim 5\times 10^{-3}$ of the peak flux density (Fig.5). In addition, the instrumental LSF
scatters only $\sim 1$\% of the line flux into broad wings, while the observed
broad wings flux in \ha\ amounts to 34\% of the line flux, again indicating that the
exponential \ha\ wings cannot be an instrumental effect.

Since the [O~III]~$\lambda 5007$ line shows no scattering wings, the scattering 
wings of \ha\ must originate in gas at radii smaller
than the NLR, most plausibly within the photoionized gas which produces the BLR
emission lines. An optical depth $\tau_e=0.34$ implies an electron column of
$5.1\times 10^{23}$~cm$^{-2}$, or equivalently a minimum total H column of about
$4.6\times 10^{23}$~cm$^{-2}$. Photoionized gas typically has an ionized,
H~II, surface
layer with a column of $\Sigma_{\rm ion} \approx 10^{23}U$~cm$^{-2}$, where
$U\equiv n_\gamma/n_e$ is the ionization parameter, and $n_\gamma$ and $n_e$
are (respectively) the H ionizing photon and the electron number densities.
To yield the required $\tau_e$ within this surface layer requires $U\sim 5$.
This value of $U$ is much larger than the typical $U\sim 0.1$ at the BLR,
and it implies $T_e\simeq 5\times 10^4$~K within the H~II layer. Such a high
$T_e$ is clearly excluded by our measurement of $T_e=1.14\times 10^4$~K within
the \ha\ scattering layer. However, below the surface H~II layer lies an extended layer
of partially ionized gas, where the lower ionization lines (Mg~II, Fe~II), and
most of the Balmer lines originate. The \ha\ scattering wings may therefore
originate in this deeper layer.

The ionization structure of the partially ionized layer cannot be deduced by
simple arguments, as made above for the H~II layer. We therefore use the
photoionization code CLOUDY (Ferland
et al. 1998) to calculate $\tau_e$, as a function of $U$ and total column
$\Sigma_{\rm H}$, for models with a significant H~I column. The partially ionized
layer is heated by the harder, more
penetrating radiation. We assume a spectral energy distribution (SED) having
an energy spectral slope of $-1$ from 1~eV to 10~eV, $-1.5$
from 10~eV to 1~keV, $-0.2$ from 1~keV to 40~keV, and a cutoff above
40~keV, which is consistent with the SED measured for NGC~4395 by Moran et al. (1999).
This X-ray hard SED produces about twice the $\tau_e$ obtained for the softer
Mathewes \& Ferland (1987) SED,
commonly used for photoionization calculations.
The observed $\tau_e=0.34$ is obtained for $\Sigma_{\rm H}=10^{24}$~cm$^{-2}$
and $U=0.3$, well within the plausible range for the BLR. The CLOUDY model calculations
show there is a significant
temperature gradient within the electron scattering layer. The surface temperature
is $3.4\times 10^4$~K, dropping to $1.6\times 10^4$~K at a depth where $\tau_e=0.07$,
$7.2\times 10^3$~K at $\tau_e=0.19$, and $7\times 10^3$~K at $\tau_e=0.3$.
Calculating the electron scattering profile of a nonisothermal slab requires
a complete radiation transfer calculation, including the \ha\ emissivity as
a function of depth, which is well beyond the scope of this
paper. However, our isothermal electron scattering model may capture the
$\tau_e$ weighted mean temperature of the electron scattering medium, which
is rather close to the $T_e\simeq 1.1\times 10^4$~K derived above.

\subsection{Is it really electron scattering?}
Although our best fit model parameters, $T_e$ and $\tau_e$, are tightly
constrained by the observed \ha\ profile, it is not clear that this model is the
only possible interpretation. One may assume that no electron scattering is present,
and that the \ha\ gas emissivity as a function of velocity in the BLR
just happens to produce
an exponential profile. How can the electron scattering model be tested?
A clear prediction of our model is that $\tau_e$ should be tightly correlated with
the ionizing flux. As the ionizing flux increases the column of ionized gas must
increase, and the relative fraction of line flux in the scattering wings should
increase. If the BLR gas has a large column, so it is ionization bounded 
rather than matter bounded, then the relation between
$\tau_e$ and the ionizing flux should be linear. In contrast, the slope of the
exponential wings $\sigma$ should remain nearly constant. This occurs
since $T_e$ is only weakly dependent on $U$ (for $U<1$), and thus should vary
only slightly. In addition, although $\tau_e$ is expected to vary linearly
with  $U$, $\sigma$ is only weakly dependent on $\tau_e$, and would thus not be
significantly affected. The ionization structure responds on a timescale of
$\sim 4\times 10^{12}/n_e$ seconds, or $\sim 400$ seconds for the BLR, which is
much smaller than the BLR light crossing time of $\sim 1$~hour (Peterson et al. 2005).
Thus, one expects the measured $\tau_e$ and $T_e$ to respond to changes in the ionizing
continuum on timescales as short as $\sim 1$~hour. However, since the dynamical
time in the BLR of NGC~4395 is $\sim 1$~week, structural changes may occur
on this time scale, or longer, which could eliminate the correlation of $\tau_e$ and $T_e$
with the ionizing flux (e.g. due to changes in the BLR density and column density).

Indirect evidence for electron scattering in NGC~4395
is provided by the spectropolarimetry of Barth et al. (1999), who find a
wavelength independent polarization of $0.67\pm 0.03$\%. The similar low polarization
of the continuum and the broad lines suggests the scattering most likely 
occurs by a low $\tau_e$ scattering medium outside the BLR, rather than within the 
BLR gas, as suggested here
\footnote{Inspection of the 2-D spectral images of NGC~4395 obtained with the HST STIS 
suggests some spatially extended scattered nuclear light.}.
Barth et al. suggested that ``A clear sign of electron scattering would be 
a broadened \ha\ profile in Stokes flux,". The exponential \ha\ wings revealed in this
study are probably not related to the Barth et al. prediction, unless scattering off the 
BLR clouds can produce the observed continuum and line polarization (Korista \& Ferland 1998).
Spectropolarimetry at a higher S/N and higher spectral resolution is required to better
constrain the location and identity of the polarizing medium in NGC~4395. 

Indirect support for the electron-scattering-within-the-BLR interpretation
is provided by the fact that best fit values of the two free model parameters, 
$T_e$ and $\tau_e$,
can apparently be obtained using a single parameter, $U\sim 0.3$.
This value of $U$ is consistent with the range of $U$ values typically deduced
in AGN, though it is significantly higher than deduced by Kraemer et al. (1999)
for this object. Can the required $U$ and $\Sigma_{\rm H}$ be
accommodated within the small BLR of NGC~4395? To estimate the physical size of such
a BLR component we use the mean ``visit 3" flux density of
$f_{\lambda}=3\times 10^{-15}$~erg$^{-1}$~s$^{-1}$~cm$^{-2}$~\AA$^{-1}$ at 1350\AA,
taken from Table 3 in Peterson et al. (2005, note an Erratum with corrected flux), 
which matches well an
extrapolation of our ESI spectrum to the UV. At a distance of 4.3~Mpc
(Thim et al. 2004), the corresponding monochromatic luminosity is
$\lambda L_{\lambda}=1.05\times 10^{40}$~erg$^{-1}$~s$^{-1}$, corrected for
Galactic extinction of 0.175 mag at 1350\AA\ (Schlegel et al. 1998 and Table 1
in Mathis 1990). Assuming the ionizing luminosity is $\simeq 3 \lambda L_{\lambda}$
at 1350\AA, and a mean ionizing photon energy of $\simeq 2$ Rydberg, the implied
ionizing photon flux from the source is $\dot{N}\simeq 7\times 10^{50}$~s$^{-1}$.
Using $U\equiv \dot{N}/(4\pi r^2c n_e)$ we get $U\simeq 20 r_{14}^{-2}n_{10}^{-1}$,
where the distance of the emitting gas is $r=10^{14}r_{14}$~cm, and its density is
$n_e=10^{10}n_{10}$~cm$^{-3}$. The measured value of $U=0.3$ implies a distance
$r_{14}=8n_{10}^{-0.5}$. The thickness of the gas slab is
$d_{14}=\Sigma_{24}/n_{10}$~cm, and thus for $\Sigma_{24}=1$ found here we get
$d/r=0.12n_{10}^{-0.5}$. Therefore, $d/r\ltsim 0.1$ for $n_e\gtsim 10^{10}$~cm$^{-3}$,
and the implied column of \ha\ emitting gas can be accommodated within the compact
BLR of NGC~4395, for reasonable values of $n_e$.

Why then are such exponential
wings not commonly seen in other AGN?  NGC~4395 is unique among type I AGN in having extremely
narrow Balmer lines. The broad \ha\ FWHM here is just 520~km~s$^{-1}$, while the typical
\ha\ FWHM in AGN is $\sim 3000$ ~km~s$^{-1}$, with a few times broader line base. Thus, the
electron scattering effect of the BLR gas will have a negligible effect on the emission
line width in most AGN. However, one can make a clear prediction that the
exponential wings should be discernable in other objects with unusually narrow lines.
A reliable detection of such wings requires a high S/N spectrum (e.g. 10-70 here at
$1000<|v|<2500$~km~s$^{-1}$).
For example, the \hb\ line wings of NGC~4395, and the \ha\ line wings in a Keck spectrum of
POX~52 (Barth et al. 2004, kindly provided by A. Barth), are broadly consistent with
an exponential form, but the available S/N in both cases is
too low to constrain the profile shape as accurately as done here. We note in passing that
as long as $\tau_e$
remains low ($\ltsim 0.5$), electron scattering would not have a significant effect on measurements
of the line FWHM, and thus on estimates of the black hole mass (e.g. Kaspi et al.
2005).

\section{Conclusions}

High quality Keck observations of the \ha\ line in NGC~4395 reveal symmetric
exponential wings of the form $f_v\propto e^{-v/\sigma}$, with $\sigma\simeq 500$~km~s$^{-1}$.
The wings are fit well by an isothermal and isotropic electron scattering model
with $\tau_e=0.34$, and $T_e=1.14\times 10^4$K, plausibly generated
in photoionized gas with $U\simeq 0.3$. The lack of electron scattering wings for
the [O~III]~$\lambda 5007$ line, and the value of $U$ indicate that the scattering occurs
within the BLR gas where \ha\ is formed.

The electron scattering interpretation can be tested by looking for a strong
correlation between $\tau_e$ and the ionizing flux. High quality spectra of other AGN with
very narrow Balmer lines will provide a further test of the electron scattering
interpretation. If the electron scattering origin is verified, then the extended
\ha\ wings can provide us with a new direct probe of $T_e$ and
$\tau_e$ within the BLR gas, and their time variability.

\acknowledgments
I would like to thank A. Barth for kindly providing the data, A. Barth and the
referee for very useful comments, and G. Ferland for making CLOUDY publicly available.
This research was supported by
The Israel Science Foundation (grant \#1030/04), and by a grant from
the Norman and Helen Asher Space Research Institute.

\end{document}